\documentclass{webofc}
\usepackage[varg]{txfonts}   
\usepackage{epstopdf}
\usepackage{amsmath}
\usepackage{amssymb}
\DeclareMathOperator{\Tr}{Tr}
\DeclareMathSizes{10}{9}{7}{5} 
\begin{document}
\title{Lattice-Boltzmann simulation of free nematic-isotropic interfaces}
%
%

\author{\firstname{Rodrigo C. V.} \lastname{Coelho}\inst{1,2}\fnsep\thanks{\email{rcvcoelho@fc.ul.pt}} \and
        \firstname{Nuno A. M.} \lastname{Araújo}\inst{1,2} \and
        \firstname{Margarida M.} \lastname{Telo da Gama}\inst{1,2}}

\institute{Centro de Física Teórica e Computacional, Faculdade de Ciências,
Universidade de Lisboa, P-1749-016 Lisboa, Portugal.\and Departamento de Física, Faculdade de Ciências,
Universidade de Lisboa, P-1749-016 Lisboa, Portugal.}

\abstract{%
We use a hybrid method of lattice Boltzmann and finite differences to simulate flat and curved interfaces between the nematic and isotropic phases of a liquid crystal described by the Landau-de Gennes theory. For the flat interface, we measure the interfacial velocity at different temperatures around the coexistence. We show that the interface is completely static at the coexistence temperature and that the profile width is in line with the theoretical predictions. The interface is stable in a range of temperatures around coexistence and disappears when one of the two phases becomes mechanically unstable. We stabilize circular nematic domains by a shift in temperature, related to the Laplace pressure, and estimate the spurious velocities of these lattice Boltzmann simulations.
}
\maketitle
\section{Introduction}
\label{intro}

Liquid crystals are systems that flow as liquids but where the particles are partially organized in crystal-like ways~\cite{p1995physics, beris1994thermodynamics}. They are formed by elongated particles the orientations of which may point in a preferential direction. At low temperatures, these liquid crystals are in the nematic phase (orientationally ordered) while at high temperatures, they are in the isotropic phase, where the orientation of the particles is random (orientationally disordered). At the nematic-isotropic temperature, the two phases coexist and a stable interface may be observed. The standard theory that describes this static interface is based on a tensor order parameter expansion, with gradient terms that account for the order parameter variations, proposed by de Gennes~\cite{p1995physics}.  

To simulate the dynamics of liquid crystals, it is common to solve numerically the continuum equations of Beris-Edwards and Navier-Stokes~\cite{marenduzzo07steady, C9SM00859D, Doostmohammadi2016}. A hybrid method of lattice Boltzmann~\cite{kruger2016lattice} and finite differences~\cite{vesely2001computational} is used. Special care has to be taken in order to simulate interfaces due to spurious currents that arise in these numerical methods, which can lead to interfacial motion or changes in the director field~\cite{kruger2016lattice, PhysRevE.82.046708}. 

We analyze the applicability of the method of Ref.~\cite{C9SM00859D} to simulate free interfaces, which are very sensitive to these spurious numerical effects. We show that it is possible to simulate static flat interfaces at coexistence, with the profiles predicted by the Landau-de Gennes theory. We estimate the magnitude of the spurious velocities at nearly static circular interfaces stabilized by temperature shifts, related to the Laplace pressure arising from the curvature.

\subsection{Equations of motion}

To simulate the liquid crystal dynamics, we solve the Beris-Edwards equation coupled to the Navier-Stokes equation. The first describes the evolution of the order parameter $Q_{\alpha \beta} = S(n_\alpha n_\beta - \delta_{\alpha \beta}/3)$
\begin{align}
  \partial _t Q_{\alpha \beta} + u _\gamma \partial _\gamma Q_{\alpha \beta} - S_{\alpha \beta}(W_{\alpha\beta}, Q_{\alpha\beta}) = \Gamma H_{\alpha\beta} .
  \label{beris-edwards}
\end{align}
Here, $\xi$ stands for the alignment parameter, $\Gamma$ is the rotational diffusive constant and the co-rotational term is given by:  
\begin{align}
 S_{\alpha \beta} =& ( \xi D_{\alpha \gamma} + W_{\alpha \gamma})\left(Q_{\beta\gamma} + \frac{\delta_{\beta\gamma}}{3} \right) 
 + \left( Q_{\alpha\gamma}+\frac{\delta_{\alpha\gamma}}{3} \right)(\xi D_{\gamma\beta}-W_{\gamma\beta})   - 2\xi\left( Q_{\alpha\beta}+\frac{\delta_{\alpha\beta}}{3}  \right)(Q_{\gamma\epsilon} \partial _\gamma u_\epsilon),
\end{align}
where $W_{\alpha\beta} = (\partial _\beta u_\alpha - \partial _\alpha u_\beta )/2$, $D_{\alpha\beta} = (\partial _\beta u_\alpha + \partial _\alpha u_\beta )/2$. The Landau-de-Gennes free energy reads
\begin{align}
    \mathcal{F} = \int_V d^3r \Big[\frac{A_0}{2}\left( 1- \frac{\gamma}{3} \right) Q_{\alpha \beta} Q_{\alpha \beta} - \frac{A_0\gamma}{3} Q_{\alpha \beta} Q_{\beta \gamma} Q_{\gamma \alpha}  +  \frac{A_0\gamma}{4} (Q_{\alpha \beta} Q_{\alpha \beta} )^2 +\frac{K}{2} (\partial _\gamma Q_{\alpha \beta}) (\partial _\gamma Q_{\alpha \beta}) \Big],
    \label{free-energy-eq}
\end{align}
where $A_0$ is a constant, $\gamma$ is a temperature related parameter and $K$ is the elastic constant. Thus, the molecular field, $H_{\alpha\beta} = -\delta \mathcal{F}/\delta Q_{\alpha\beta} + (\delta_{\alpha\beta}/3) \Tr (\delta \mathcal{F}/\delta Q_{\gamma \epsilon} )$, is
\begin{align}
 H_{\alpha\beta} = -A_0\left( 1-\frac{\gamma}{3}\right)Q_{\alpha\beta} - A_0\gamma\left( Q_{\nu\gamma}^2\frac{\delta_{\alpha\beta}}{3} - Q_{\alpha\gamma}Q_{\beta\gamma} \right) - A_0\gamma Q_{\alpha\beta}Q_{\gamma\lambda}^2 + K\partial_\gamma\partial_\gamma Q_{\alpha\beta}.
 \label{molecular-field-eq}
\end{align}

The continuity and Navier-Stokes equations describe the evolution of the density and velocity of the fluid
\begin{align}
 \partial _t \rho + \partial_\alpha (\rho u_\alpha) =0, \quad \quad
\rho \partial _t u_\alpha + \rho u_\beta \partial_\beta u_{\alpha} = \partial _\beta \Pi_{\alpha\beta}  + \eta \partial_\beta\left( \partial_\alpha u_\beta + \partial_\beta u_\alpha \right).
\label{navier-stokes}
\end{align}
where, the stress-tensor is
\begin{align} 
 \Pi_{\alpha\beta} =& -P_0 \delta_{\alpha\beta} + 2\xi \left( Q_{\alpha\beta} +\frac{\delta_{\alpha\beta}}{3} \right)Q_{\gamma\epsilon}H_{\gamma\epsilon} - \xi H_{\alpha\gamma} \left( Q_{\gamma\beta}+\frac{\delta_{\gamma\beta}}{3} \right) - \xi \left( Q_{\alpha\gamma} +\frac{\delta_{\alpha\gamma}}{3} \right) H_{\gamma \beta} \nonumber \\ & 
 - \partial _\alpha Q_{\gamma\nu} \,\frac{\delta \mathcal{F}}{\delta (\partial_\beta Q_{\gamma\nu})} + Q_{\alpha\gamma}H_{\gamma\beta} - H_{\alpha\gamma}Q_{\gamma\beta} .
 \label{passive-pressure-eq}
\end{align}
$P_0=\rho c_s^2$ is the hydrostatic pressure. Using Eq.~\eqref{free-energy-eq}, we obtain $\delta \mathcal{F}/\delta (\partial_\beta Q_{\gamma\nu}) = K\partial_\beta Q_{\gamma \nu}$, which can be replaced in Eq~\ref{passive-pressure-eq}.

\subsection{Hybrid method}
\label{hybrid-method-sec}

Our numerical scheme solves Eq.~\eqref{beris-edwards} using finite differences and Eq.~\eqref{navier-stokes} using lattice Boltzmann. Both methods are solved using the same grid (spatial discretization) and the same time step.

\subsubsection{Lattice Boltzmann}

The lattice Boltzmann method is a numerical technique which solves the Boltzmann equation and recovers the Navier-Stokes equation in the macroscopic limit. The space is discretized in a regular grid and the velocity space is discretized according to the lattice (Gaussian quadrature). Here, we use the D3Q19 lattice, which has 19 velocity vectors $\mathbf{c}_i$ isotropically distributed in three dimensions. The discrete version of the Boltzmann equation is implemented in the method
\begin{align}
  &f_i(\mathbf{x}+\mathbf{c}_i \Delta t, t+\Delta t) - f_i(\mathbf{x}, t) = \left( \partial f_i / \partial t\right)_{\text{coll}} + S_i,
\end{align}
where $f_i$ is the distribution function corresponding to the $\mathbf{c}_i$ vector. The simplest collision operator is the Bhatnagar–Gross–Krook (BGK) one, which assumes that the full-equilibrium distribution $f_i$ relaxes to the equilibrium $f_i^{eq}$ with a characteristic relaxation time $\tau$: $\left( \partial f_i / \partial t\right)_{\text{coll}} = -(\Delta t/\tau)(f-f^{eq}) $. The multi-relaxation time operator, used in this work, is a generalization of the BGK: $\left( \partial f_i / \partial t\right)_{\text{coll}}=\mathbf{M}^{-1}\mathbf{R}\mathbf{M} [ f_i(\mathbf{x}, t) - f_i^{eq}(\mathbf{x}, t) ]\Delta t$, where the matrix $\mathbf{M}$ transforms from the distribution function space to the hydrodynamic moments space and $\mathbf{R}$ is the relaxation matrix. It assumes that the hydrodynamic moments may relax with different relaxation times. This approach is known to improve the accuracy and stability in many problems~\cite{kruger2016lattice}. The equilibrium distribution is the second order expansion in Hermite polynomials of the Maxwell-Boltzmann distribution~\cite{COELHO2018144}:
\begin{align}
  f^{eq}_i = \rho w_i \left[ 1+ \frac{\mathbf{c}_i\cdot\mathbf{u}}{c_s^2} + \frac{(\mathbf{c}_i\cdot\mathbf{u})^2}{2c_s^4} - \frac{\mathbf{u}^2}{2c_s^2}  \right],
\end{align}
where $c_s$ is the speed of sound in the given lattice. For the D3Q19, it is $c_s=1/\sqrt{3}$. The density and velocity fields are calculated as follows:
\begin{align}
  \rho = \sum _i f_i, \quad \rho \mathbf{u} = \sum _i \mathbf{c}_i f_i + \frac{\mathbf{F}_i \Delta t}{2} .
  \label{density-vel-eq}
\end{align}
The source term, for the BGK operator, reads: $S_i = w _i\left(1-\frac{1}{2\tau} \right) \left[ \frac{\mathbf{c}_i}{c_s^2}\left(1+  \frac{\mathbf{c}_i\cdot \mathbf{u}}{c_s^2}\right) - \frac{\mathbf{u}}{c_s^2} \right]\cdot \mathbf{F}$. For the MRT operator, this source term must be relaxed in the moments space as the distribution function (see Refs.~\cite{C9SM00859D, kruger2016lattice} for more details of the MRT method). The force is calculated using the stress tensor of Eq.~\eqref{passive-pressure-eq}: $F_\alpha = \partial_\beta(\Pi_{\alpha\beta}+P_0\delta_{\alpha\beta})$.

\subsubsection{Finite differences}

The Eq.~\eqref{beris-edwards} is solved explicitly using a predictor-corrector finite difference method. All the differences are second order accurate. For instance, the first derivative of a vector $V$ in the $x$-direction  is: $  \partial_x V(x)= [V(x+\Delta x) - V(x-\Delta x)] /(2\Delta x) + \mathcal{O}(\Delta x^2)$. To calculate the time evolution of $Q_{\alpha\beta}$, we follow the following steps: 1) Calculate the time derivative, $(\partial Q_{\alpha\beta}/\partial t)^\prime$, using Eq.~\eqref{beris-edwards} with the current $Q_{\alpha\beta}$; 2) Calculate the predictor: $Q^P_{\alpha\beta} = Q_{\alpha\beta} + \Delta t (\partial Q_{\alpha\beta}/\partial t)^\prime $; 3) Calculate the time derivative using the predictor, $(\partial Q_{\alpha\beta}^P/\partial t)$, with Eq.~\eqref{beris-edwards}; 4) Calculate the corrected derivative: $\frac{\partial Q_{\alpha\beta}}{\partial t} = \frac{1}{2}\left[ \left(\frac{\partial Q_{\alpha\beta}}{\partial t}\right)^\prime +\left(\frac{\partial Q_{\alpha\beta}^P}{\partial t} \right)\right]$;
5) Calculate the corrector: $Q_{\alpha\beta} (t+\Delta t) = Q_{\alpha\beta} (t) + \Delta t\frac{\partial Q_{\alpha\beta}}{\partial t}$; 6) Return to step 1. Notice that the velocity used in Eq.~\eqref{beris-edwards} is the actual one (and not the lattice Boltzmann one: $\rho \mathbf{u}_{LB} = \sum _i \mathbf{c}_i f_i$), given by Eq.~\eqref{density-vel-eq}, which is second order accurate.

\section{Free interface}

In this section, we show the results from numerical simulations of unconfined nematic-isotropic interfaces. First, we consider a flat interface in two dimensions and then we simulate a circular interface. In the simulations, we apply periodic boundary conditions in both directions. The initial velocity is set to zero and the density to $\rho=1$ everywhere. Other parameters are: $\tau=1.5$, $K=0.04$, $\xi=0.7$, $A_0=0.1$ and $\Gamma=0.34$. Our results are given in lattice units: the distance between nodes is $\Delta x = 1$ and the time step is $\Delta t = 1$.

\subsection{Flat interface}
\begin{figure}[t]
\center
\includegraphics[width=1.0\linewidth]{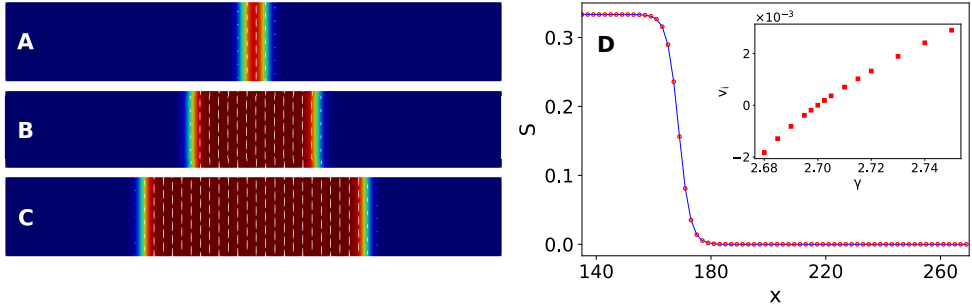}
\caption{Order parameter for flat interfaces at different temperatures and at time $t=350000$: A) $\gamma = 2.699$, B) $\gamma=2.7$, C) $\gamma=2.701$. The velocity field is below the machine precision everywhere for $\gamma=2.7$. D) Scalar order parameter at $y=L_Y/2$. The solid line is the analytical expression given by Eq.~\eqref{x-interface-eq} for $S_n=1/3$ and $\lambda=2$ and the circles are the simulated results. The inset shows the interface velocity at different temperatures. At coexistence $\gamma=2.7$, the interface velocity is below the machine precision.}
\label{gamma-fig}
\end{figure}

In order to apply periodic conditions, we initialize our system as follows: the nematic phase, with directors in the vertical direction, is in the center of the domain with isotropic phase elsewhere. Thus, we simulate two interfaces, but we focus our analysis on the right interface. At the coexistence temperature ($\gamma_{NI}=2.7$) the interface is completely static and there are no spurious velocities. Eq.~\eqref{free-energy-eq} allows an interface solution of the form~\cite{MESQUITA1998}
\begin{align}
 S = \frac{S_n}{2}\left[  1-\tanh\left(  \frac{x-x_{i}}{2 \lambda}   \right )   \right],
 \label{x-interface-eq}
\end{align}
where $S_n$ is the nematic order parameter and $\lambda=\sqrt{27 K/A_0 \gamma}$ is the correlation length. At $\gamma=2.7$, $\lambda=2$ and $S_n=1/3$. Fig. \ref{gamma-fig} D shows that the simulation reproduces accurately the interfacial profile. 

Next, we change the temperature slightly away from coexistence. In this case, the interface moves with constant velocity, which can be to the left (negative) or to the right (positive). Figs.~\ref{gamma-fig} A and C, show the interfaces at two different temperatures after an interval of time. For $\gamma < \gamma_{NI}$ the domain of the nematic phase shrinks and for $\gamma > \gamma_{NI}$ the nematic domain expands. In the inset of Fig.~\ref{gamma-fig}D, we plot the interface velocity at different temperatures. It is approximately linear close to $\gamma_{NI}$ but is deviates from the linear behavior as the shift from coexistence increases. Above a certain shift from $\gamma_{NI}$ one of the two phases becomes unstable and the entire domain becomes nematic or isotropic in the absence of interfacial propagation. Below $\gamma_<=2.65$ the system becomes isotropic and above $\gamma_>=3.1$ it becomes nematic.

\subsection{Circular interface and spurious velocities}

\begin{figure}[t]
\center
\includegraphics[width=1.0\linewidth]{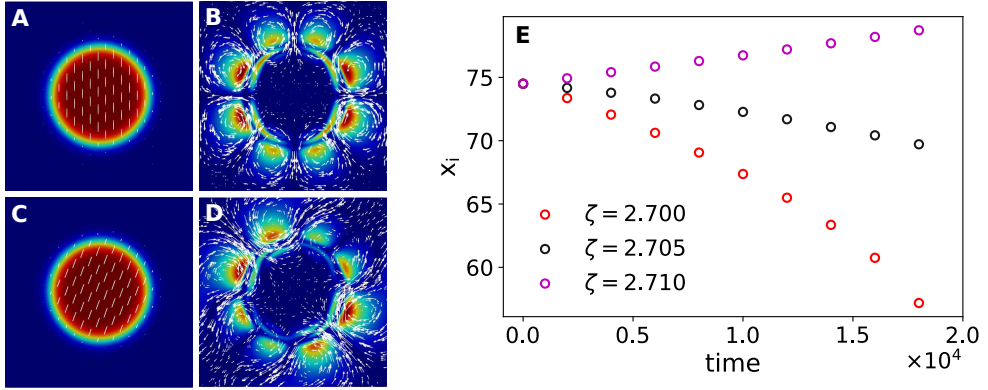}
\caption{Order and velocity fields of an almost static circular interface at two different director orientations. The temperature, $\gamma = 2.707681$ was chosen to minimize the interfacial velocity and the screenshots were taken at $t=35000$. The system size is $100\times 100$, the radius of the circle is $R=25$. A) Order parameter (colors) and directors (lines) for $n_y=1$ and $n_x=0$. B) Vorticity field (colors) and direction of the velocity field (arrows). C and D depict the same fields as A and B, but for $n_y = \cos(20^{\circ})$ and $n_x = \sin(20^{\circ})$. In this color map, the red and blue represent the maximum and minimum values of the field. The maximum velocity (norm) in B is $u_{max}=8.97\times 10^{-7}$ and $u_{max}=1.13\times 10^{-6}$ in D. E) Interface velocity at three different temperatures.}
\label{circle-fig}
\end{figure}

Here we estimate the spurious velocities generated at a circular interface. The spurious velocities arise at curved interfaces as a result of the discretization of the velocity space. However, it is difficult to distinguish the spurious velocities from the physical ones in this multiphase model, because curved interfaces are usually unstable. Fig.~\ref{circle-fig}E shows the interface position at three different temperatures. In order to identify the spurious velocities, we iteratively search for the temperature where the interface velociy is of the order of the spurious ones. We found that for $\gamma = 2.707681$ the interface velocity at $t=35000$ is approximately $u_i\approx 1.89\times 10^{-7}$. Fig.~\ref{circle-fig}A and B show the oder parameter and velocity field of this almost static interface. From the vorticity field, we notice that the there are vortices between the directions of the velocity vectors of the D3Q19 lattice, which is an indication that these are spurious currents. In Fig.~\ref{circle-fig} D we plot the velocity field for the same system rotated by 20$^\circ$. One notes that the intensity of the vortices changed due to the physical component of these velocities (the small interface velocity), but the positions of the vortices are the same confirming that they are mostly due to the velocity discretization. From this analysis, we estimate that the spurious velocities are around $10^{-6}$ in lattice units for this setup. If necessary, one could reduce the spurious velocities by increasing the isotropy of the lattice~\cite{PhysRevE.73.047701} or using more advanced collision operators~\cite{2019arXiv190412948C}.

\section{Conclusions}

We examined the applicability of the hybrid method of lattice Boltzmann and finite differences for the simulation of nematic-isotropic interfaces. The shape and width of the interface is correctly reproduced and the free interface is static at the coexistence temperature. At temperatures slightly different from the coexistence, the interface moves with constant velocity that depends on the temperature. The interface exists only for a certain range of temperatures: for temperatures much lower or higher than the coexistence, one of the two phases becomes unstable and the interface disappears without propagation. The circular interfaces are not static at the coexistence temperature. Thus, we stabilize them by a temperature shift in a way that the interface velocity is comparable to the spurious velocities. This allows us to visualize the profile of theses spurious currents and to estimate their magnitude.

\section*{Acknowledgements}

We acknowledge financial support from the Portuguese Foundation for Science and Technology (FCT) under the contracts: PTDC/FIS-MAC/28146/2017 (LISBOA-01-0145-FEDER-028146) and UID/FIS/00618/2019. Margarida Telo da Gama (MTG)
would like to thank the Isaac Newton Institute for Mathematical Sciences for support and hospitality during the program ”The mathematical design of new materials” where most of this work was carried out. This program was supported by EP-SRC 
Grant Number: EP/R014604/1. MTG participation in the program was supported in part by a Simons Foundation Fellowship.

%
%
\bibliography{rsc} 

\begin{thebibliography}{12}

\bibitem{p1995physics}
P.G. de~Gennes, J.~Prost, \emph{The Physics of Liquid Crystals}, International
  Series of Monographs on Physics (Clarendon Press, 1995), ISBN 9780198517856

\bibitem{beris1994thermodynamics}
A.~Beris, B.~Edwards, \emph{Thermodynamics of Flowing Systems: with Internal
  Microstructure}, Oxford Engineering Science Series (Oxford University Press,
  1994), ISBN 9780195344882

\bibitem{marenduzzo07steady}
D.~Marenduzzo, E.~Orlandini, M.E. Cates, J.M. Yeomans, Physical Review E
  \textbf{76}, 031921 (2007)

\bibitem{C9SM00859D}
R.C.V. Coelho, N.A.M. Araújo, M.M. Telo~da Gama, Soft Matter  (2019), accepted
  manuscript - DOI:10.1039/C9SM00859D

\bibitem{Doostmohammadi2016}
A.~Doostmohammadi, M.F. Adamer, S.P. Thampi, J.M. Yeomans, Nature
  Communications \textbf{7}, 10557 (2016)

\bibitem{kruger2016lattice}
T.~Kr\"uger, H.~Kusumaatmaja, A.~Kuzmin, O.~Shardt, G.~Silva, E.M. Viggen,
  \emph{The Lattice Boltzmann Method - Principles and Practice} (Springer
  International Publishing, 2016), ISBN 978-3-319-44647-9

\bibitem{vesely2001computational}
F.~Vesely, \emph{Computational Physics: An Introduction} (Springer US, 2001),
  ISBN 9780306466311,
  \urlstyle{tt}\url{https://www.springer.com/gp/book/9780306466311}

\bibitem{PhysRevE.82.046708}
Z.~Yu, L.S. Fan, Physical Review E \textbf{82}, 046708 (2010)

\bibitem{COELHO2018144}
R.C.V. Coelho, M.M. Doria, Computers \& Fluids \textbf{165}, 144  (2018)

\bibitem{MESQUITA1998}
O.N. Mesquita, {Brazilian Journal of Physics} \textbf{28}, 04 (1998)

\bibitem{PhysRevE.73.047701}
X.~Shan, Physical Review E \textbf{73}, 047701 (2006)

\bibitem{2019arXiv190412948C}
C.~{Coreixas}, B.~{Chopard}, J.~{Latt}, arXiv e-prints arXiv:1904.12948 (2019),
  \texttt{1904.12948}

\end{thebibliography}
\bibliographystyle{rsc} 

%

\end{document}